\documentstyle[12pt]{article}

\newcommand{\nc}{\newcommand}
\nc{\be}{\begin{equation}}
\nc{\ee}{\end{equation}}
\nc{\bea}{\begin{eqnarray}}
\nc{\eea}{\end{eqnarray}}
\nc{\lsm}{L$\sigma$M}
\nc{\lf}{\left}
\nc{\rgt}{\right}
\nc{\refa}[1]{(\ref{#1})}

\begin{document}

\title{Regularizing the quark-level linear $\sigma$ model.}

\author{R. Delbourgo, \makebox[1cm]{}
	M.D. Scadron\thanks{Permanent address: 
	Physics Department, University of Arizona, Tucson, Az. 85721 USA.}\\
	Physics Department\\
	University of Tasmania\\
	Hobart, Tasmania, 7005
	\vspace{0.5cm}\\
	and
	\vspace{0.5cm}\\
	A.A. Rawlinson\\
	School of Physics\\
	University of Melbourne\\
	Parkville, Australia, 3052.}

\date{\today}
\maketitle

\begin{abstract}\noindent 
We show that the finite difference, $-i\pi^2 m^2$, between quadratic
and logarithmic divergent integrals 
$\int d^4p\left[m^2(p^2-m^2)^{-2}-(p^2-m^2)^{-1}\right]$, as encountered 
in the linear $\sigma$ model, is in fact regularization independent.

\end{abstract}

\vspace{0.5cm}


\newpage

Given that the scalar meson $\sigma$ (now called the $f_0(400-900)$ 
in~\cite{torn}) will shortly be reinstated in the 1996 particle data group
tables, but perhaps with a slightly extended mass range, we believe it
important for theorists to reconsider seriously the original $SU(2)$ chiral 
linear $\sigma$ model field theory~\cite{gell}. In a recent letter~\cite{ds} 
it was shown that a dimensionally regularized quark-level linear $\sigma$
model (\lsm) not only dynamically generates the spontaneously broken
Gell-Mann-L\'evy interacting Lagrangian density~\cite{gell}
\be
{\cal L}^{int}_{L\sigma M} = g\bar{\psi}\lf(\sigma'+i\gamma_5\vec{\tau}\cdot
\vec{\pi}\rgt)\psi+g'\sigma'\lf(\sigma'^2+\vec{\pi}^2\rgt)-\lambda
\lf(\sigma'^2+\vec{\pi}^2\rgt)^2/4
\label{eq1}
\ee
with the chiral-limiting meson-quark and meson-meson couplings
\be
g={m_q\over f_\pi}, \makebox[2cm]{} g'={m_\sigma^2\over 2 f_\pi} = 
\lambda f_\pi,
\label{eq2}
\ee
but that the \lsm ~parameters satisfy the one-loop order relations
\be
m_\sigma=2 m_q, \makebox[2cm]{} g={2\pi\over\sqrt{N_c}}.
\label{eq3}
\ee

Then for color number $N_c=3$ and chiral limiting $f_\pi\approx 90$ MeV, 
\refa{eq2} and \refa{eq3} require
\bea
&m_q=f_\pi{2\pi\over\sqrt{3}}\approx 325\,{\rm MeV},&\makebox[2cm]{}
m_\sigma=2 m_q\approx 650\,{\rm MeV},\label{eq4}\\
&g={2\pi\over\sqrt{3}}=3.6276, &\makebox[2cm]{} g_{\pi NN}=
3 g_A g\approx 13.68,
\label{eq5}
\eea
for the measured value \cite{pdg} $g_A\approx 1.2573$. Since the predicted
$\pi NN$ coupling in~\refa{eq5} is near the phenomenologically 
determined~\cite{bugg} $g_{\pi NN}=13.40\pm0.08$ and~\refa{eq4} predicts
a reasonable constituent quark mass $m_q$ near $M_N/3$, along with a
Nambu-Jona-Lasinio~\cite{njl} scalar~$\sigma$ mass which is the 
average~\cite{comment} of the non-strange empirical $\sigma$ mass
extracted in~\cite{torn}, the dynamically generated~\cite{ds} \lsm ~appears 
to reflect reality.

In this note we show that the crucial new relations of \refa{eq3}
are in fact {\em regularization scheme independent}. Specifically
they hold not only for dimensional regularization, but also
analytic regularization~\cite{del}, a symmetrical approach to
generalized functions~\cite{lod}, and for Pauli-Villars regularization.

We begin by quickly reviewing the dynamical generation of the \lsm 
~Lagrangian~\refa{eq1} and chiral couplings~\refa{eq2} starting from the more
basic chiral quark model~(CQM) massless Lagrangian
\be
{\cal L}_{CQM}=\bar{\psi}\lf[i{\not \!\partial}+g\lf(\sigma+i\gamma_5
	\vec{\tau}\cdot\vec{\pi}\rgt)\rgt]\psi+
	\lf[\lf(\partial\sigma\rgt)^2+\lf(\partial\vec{\pi}\rgt)^2\rgt]/2.
\label{eq6} 
\ee
Here the bare quark, $\sigma$ and pion masses are zero in~\refa{eq6};
the quark and $\sigma$ masses must be dynamically generated from the 
meson-quark chiral driving term 
\[g\bar{\psi}\lf(\sigma+i\gamma_5\vec{\tau}\cdot
\vec{\pi}\rgt)\psi, 
\]
while the pion remains massless due to the underlying
conservation of the axial current $\partial^\mu \vec{A}_\mu=0$, which
in turn ensures the quark-level Goldberger-Treiman relation~(GTR)
$f_\pi g=m_q$ in \refa{eq2}.

The pion decay constant $f_\pi$ and quark mass $m_q$ are simultaneously
nonperturbatively generated in the spirit of NJL gap equations 
$\delta f_\pi=f_\pi$ and $\delta m_q=m_q$ via the quark loops in Figure~1.
Since there are no mass terms in the CQM Lagrangian~\refa{eq6}, the physical
masses $m_q$ and $m_\sigma$ in Fig.~1 equal the counterterm masses. Due to
the GTR $m_q=f_\pi g$, Fig.~1a generates the chiral-limiting log-divergent
gap equation (with $\bar{d}^4p=d^4p(2\pi)^{-4}$ and $N_c$ quark colors)
\be
1=-4iN_cg^2\int{\bar{d}^4p\over (p^2-m_q^2)^2},
\label{eq7}
\ee
which is a compositeness condition in the context of~\cite{salam}.
On the other hand,
Fig.~1b with two quark flavors generates~\cite{commenta} the counterterm 
$m_q$ (quadratically divergent) mass gap
\be
m_q={-8iN_c g^2\over m_\sigma^2}\int{\bar{d}^4p\; m_q\over p^2-m_q^2}.
\label{eq8}
\ee

The blending of the logarithmic and quadratic divergent loop integrals in 
\refa{eq7}~and~\refa{eq8} is where~\cite{ds} specialized to the dimensional
regularization scheme~(dim. reg.), leading to the lemma in the limit of
$2l\rightarrow 4$ dimensions,
\be
\int\bar{d}^4p\lf[{m^2\over \lf(p^2-m^2\rgt)^2}-{1\over\lf(p^2-m^2\rgt)}\rgt]
=\lim_{l\rightarrow 2}{i m^{2l-2}\over(4\pi)^l}\lf[\Gamma(2-l)+\Gamma(1-l)\rgt]
={-i m^2\over(4\pi)^2}.
\label{eq9}
\ee
Since this lemma~\refa{eq9} was derived from dimensional continuation, one
might suspect that it is specific to that regularization and is 
therefore subject to criticism. We now demonstrate that the result~\refa{eq9}
holds more generally.

First we reformulate the left hand side~(l.h.s) of~\refa{eq9} with the aid 
of the identity
\be
I = \int\bar{d}^4p\lf[{m^2\over\lf(p^2-m^2\rgt)^2}-{1\over p^2-m^2}\rgt]
=\lf(m^2{d\over dm^2}-1\rgt)\int{\bar{d}^4p\over p^2-m^2}.
\label{eq10}
\ee
The r.h.s. of~\refa{eq10} is more amenable to alternative regularization 
schemes than is the l.h.s. of~\refa{eq9} or~\refa{eq10}. Specifically the 
analytic regularization scheme uses (eq. 3.30 of~\cite{del}) non-integral
$\Sigma$,
\be
-i(4\pi)^2\int{\bar{d}^4p\over\lf(p^2-m^2\rgt)^\Sigma}
={(-1)^{-\Sigma}\Gamma(\Sigma-2)\over \Gamma(\Sigma)(m^2)^{\Sigma-2}}
={(-m^2)^{2-\Sigma}\over (\Sigma-1)(\Sigma-2)}.
\label{eq11}
\ee
Inserting \refa{eq11} into the r.h.s. of \refa{eq10} in the limit $\Sigma
\rightarrow 1$ then yields
\bea
& &\makebox[-1cm]{}
-i(4\pi)^2\lf(m^2{d\over dm^2}-1\rgt)\int{{\bar d}^4p\over
\lf(p^2-m^2\rgt)^\Sigma}\nonumber\\ 
&=&\lf[{2-\Sigma\over(\Sigma-1)(\Sigma-2)}-{1\over(\Sigma-1)(\Sigma-2)}\rgt]
(-m^2)^{2-\Sigma}\rightarrow-m^2.
\label{eq12}
\eea
Substituting \refa{eq12} back into \refa{eq10} then gives for the analytic
regularization
\be
I = \int\bar{d}^4p\lf[{m^2\over(p^2-m^2)^2}-{1\over p^2-m^2}\rgt]\rightarrow
{-i m^2\over(4\pi)^2},
\label{eq13}
\ee
which is precisely the result of the dim. reg. lemma~\refa{eq9}; 
$\zeta-$function regularization is essentially equivalent to that. 

To verify that the scheme-independent relation~\refa{eq10} also reproduces
the original dimensional regularization lemma~\refa{eq9}, one replaces the 
analytic regularization scheme integral~\refa{eq11} by~\cite{del}
\be
-i(4\pi)^l\int{\bar{d}^{2l}p\over p^2-m^2}=-\Gamma(1-l)(m^2)^{l-1}
\label{eq14}
\ee
in $2l$ dimensions. Using the identity $\Gamma(1-l)=\Gamma(2-l)/(1-l)$ 
in~\refa{eq14} and substituting the latter into the r.h.s. of~\refa{eq10}, 
one finds as $l\rightarrow 2$,
\be
-i(4\pi)^l\lf(m^2{d\over dm^2}-1\rgt)\int{{\bar d}^{2l}p\over p^2-m^2}
=-\Gamma(2-l){\lf[(l-1)-1\rgt]\over 1-l}(m^2)^{l-1}
\rightarrow -m^2.
\label{eq15}
\ee
Since the r.h.s. of \refa{eq15} is the same as the r.h.s. of~\refa{eq12},
the dim. reg. lemma~\refa{eq9} is clearly recovered, which is hardly 
surprising.

Because both the dimensional and analytic regularization schemes involve
$\Gamma-$functions, we should also consider alternative regularization
schemes not containing them. Specifically we study the symmetric generalized
function scheme advocated by Lodder~\cite{lod} involving natural
logarithmic functions:
\be
-i(4\pi)^2\int{\bar{d}^4p\over p^2-m^2}=-m^2\lf(\ln\lf({m^2\over M^2}\rgt)+
C\rgt).
\label{eq16}
\ee
Here $C$ is an indeterminate constant and $M$ is some mass scale.
Then the analogue of~\refa{eq12} and~\refa{eq15} is
\bea
& &\makebox[-1cm]{}
-i(4\pi)^2\lf(m^2{d\over dm^2}-1\rgt)\int{{\bar d}^4p\over(p^2-m^2)}
\nonumber\\ 
&=&-m^2\lf(\ln\lf({m^2\over M^2}\rgt)+C\rgt)-m^2+
m^2\lf(\ln\lf({m^2\over M^2}\rgt)+C\rgt)=-m^2,
\label{eq17}
\eea
again giving~\refa{eq13}. 

Finally, turning to Pauli-Villars regularization,
the l.h.s. of (9) or (10) can be alternatively written in the form
\be
I = \int\frac{\bar{d}^4p}{p^2}\left[\frac{m^4}{(p^2-m^2)^2} - 1]\right].
\label{eq18}
\ee
Next introduce an ultraviolet cut-off $\Lambda$ and sum over massive
fermions (masses $M_j$) with probabilities $c_j$. The integral $I$
thereby reduces to
\be
I = \sum_j ic_j(\Lambda^2 - M_j^2)/(4\pi)^2.
\label{eq19}
\ee
Applying the Pauli-Villars sum rules,~\cite{del,IZ} 
$\sum c_j = 0, \sum c_jM_j^2 = m^2$
(which eliminates the quadratic divergence or massless tadpole), we remain
once again with $I=-im^2/(4\pi)^2$, namely the r.h.s. of (9). We contend that 
any other reasonable regularization scheme will lead to the same final 
result~\refa{eq9}.

Returning to the Fig.~1b counterterm mass gap equation~\refa{eq8}, we cancel
out the quark mass $m_q$ from both sides of~\refa{eq8} to write
\be
m_\sigma^2=-8iN_c g^2\int{\bar{d}^4p\over p^2-m_q^2}.
\label{eq20}
\ee
Subtracting the quadratic divergent integral in~\refa{eq20} from the 
log-divergent mass gap integral of~\refa{eq7} weighted by $2m_q^2$, the
regularization independent lemma~\refa{eq9} then leads to
\be
m_\sigma^2=2m_q^2\lf[1+{g^2 N_c\over 4\pi^2}\rgt].
\label{eq21}
\ee
Moreover, the quark bubble plus tadpole graphs of Fig.~2 generate the 
counterterm $\sigma$ mass (squared) from the CQM Lagrangian~\refa{eq6}
giving~\cite{ds}
\be
m_\sigma^2=16i N_c g^2\lf[\int{m_q^2\; \bar{d}^4p\over(p^2-m_q^2)^2}-
\int{\bar{d}^4p\over(p^2-m_q^2)}\rgt]={N_c g^2 m_q^2\over \pi^2},
\label{eq22}
\ee
by virtue of the same lemma~\refa{eq9}. Solving the two equations~\refa{eq21}
and~\refa{eq22} in terms of the two parameters $m_q^2$ and $N_c g^2$, one
obtains the two key \lsm~relations in~\refa{eq3}, with resulting physical
scales given in~\refa{eq4} and~\refa{eq5}.

To reaffirm the NJL relation $m_\sigma=2 m_q$ in this \lsm~context and to
test the consistency of the {\em counterterm} mass relations~\refa{eq8}
or~\refa{eq21} and~\refa{eq22}, we express the integral version of~\refa{eq22}
as
\be
m_\sigma^2=-4m_q^2+2 m_\sigma^2.
\label{eq23}
\ee
Here the first log-divergent integral in~\refa{eq22} is replaced by $-4m_q^2$
in~\refa{eq23} using the log-divergent gap equation~\refa{eq7}, and the 
second quadratic-divergent integral in~\refa{eq22} is replaced by the 
counterterm $2 m_\sigma^2$ in~\refa{eq23} using~\refa{eq8} or~\refa{eq21}.
The solution of~\refa{eq23} is trivially $m_\sigma=2 m_q$, as anticipated.

\vspace{.3in}

\noindent {\bf Acknowledgements.}

This research was partially supported by the Australian Research Council.
M.D.S. appreciates hospitality of the University of Melbourne and the 
University of Tasmania. He also is grateful to V. Elias and V.A. Miransky
for incisive remarks which motivated this investigation.

\vspace{035in}

\noindent
{\bf Figure Captions}
\vspace{.2in}

\noindent Fig. 1(a) Quark loop determination of the decay constant $f_\pi$.

\noindent Fig. 1(b) Mass gap quark tadpole.

\noindent Fig. 2 Quark loop induced ($\sigma$ mass)$^2$.


\begin{thebibliography}{99}
\bibitem{comment} While the $\sigma$ has mass range $400-900$ MeV
in~\cite{torn}, the \lsm~width is 
$\Gamma(\sigma\pi\pi)={3\over 2}q\lf|m_\sigma^2/f_\pi\rgt|^2/8\pi m_\sigma^2
\approx 700$ MeV for the average $\sigma$ mass $m_\sigma=650$ MeV and is as 
wide as the mass. This is the main reason why the $\sigma$ took so long to
establish.
\bibitem{commenta} Disregarding the propagator $\sigma$ mass when evaluating
the dominant log-divergent behaviour, the quark self-energy amplitudes
for the sum of $\vec{\pi}-$ and $\sigma-$ mediated loops sum to
$$
\Sigma\lf(p\rgt)=i g^2\int {{\bar d}^4k\over k^2\lf[\lf(p+k\rgt)^2-m_q^2\rgt]}
	\lf[-3\gamma_5\lf({\not \! p}+{\not \! k}+m_q\rgt)\gamma_5
	+\lf({\not \! p}+{\not \! k}\rgt)+m_q\rgt].
$$
Introducing Feynman parameter $x$ and shifting the loop 
$4-$momentum as $k\rightarrow k-p x$, the above loop 
integral vanishes on the quark mass-shell
$$\Sigma\lf(\not\! p=m_q,p^2=m_q^2\rgt)\propto\int_0^1 dx\lf[-2\lf(1-x\rgt)
+1\rgt]=0.
$$
For this reason we have dropped the standard $\vec{\pi},\sigma$ 
contributions to the quark self-energy; they cancel to leading order by
chiral symmetry.
\end{thebibliography}
\end{document}